\documentstyle[11pt]{article}

\def\roughly#1{\mathrel{\raise.3ex\hbox{$#1$\kern-.75em%
\lower1ex\hbox{$\sim$}}}}

\def\lsim{\roughly<}

\def\be{\begin{eqnarray}}
\def\ee{\end{eqnarray}}

\def\ben{\begin{enumerate}}
\def\een{\end{enumerate}}
\def\beitem{\begin{itemize}}
\def\eitem{\end{itemize}}

\newcommand{\beq}{\begin{eqnarray}}
\newcommand{\eeq}{\end{eqnarray}}

\def\bi{\begin{itemize}}
\def\ei{\end{itemize}}
\def\ie{{\it i.e}}
\def\eg{{\it e.g.}}
\def\etal{{\it et al}}

\long\def\beginomit#1\endomit{}
\def\np{{Nucl. Phys.}}

\def\pl{Phys. Lett.}

\begin{document}

\begin{titlepage}\begin{center}

\hfill{HYUPT-94/15}

\hfill{hep-ph/9411416}

\vskip 0.4in
{\Large\bf RENORMALIZATION-GROUP FLOW ANALYSIS}
\vskip 0.1cm
{\Large\bf OF MESON CONDENSATIONS IN DENSE MATTER}
\vskip 1.2in
{\large  Hyun Kyu Lee$^a$, Mannque Rho$^b$ and Sang-Jin Sin$^a$}\\
\vskip 0.1in
{\large a) \it Department of Physics, Hanyang University,} \\
{\large \it Seoul, Korea}\\
{\large b) \it Service de Physique Th\'{e}orique, CEA  Saclay}\\
{\large\it 91191 Gif-sur-Yvette Cedex, France}\\
\vskip .6in
\centerline{November 1994}
\vskip .6in

{\bf ABSTRACT}\\ \vskip 0.1in
\begin{quotation}
\noindent We present a renormalization-group (RG) flow argument
for s-wave kaon condensation in dense nuclear-star matter predicted
in chiral perturbation theory. It is shown that it is the {\it relevant}
mass term together with  {\it any}  attractive interaction
for the kaon in medium that triggers the instability.
We show that a saddle point of multi-dimensional RG flow
can imply a phase transition. Pion condensation is also analyzed
along the same line of reasoning.
\end{quotation}
\end{center}\end{titlepage}


The idea of renormalization group (RG)
has been used extensively  both in condensed
matter physics and particle physics, especially for the critical
phenomena  and
various situations involving scaling behavior.
Recently, Shankar\cite{shankar} and Polchinski\cite{polchinski1}
 showed that one can use the RG idea even
for the phenomena including a scale, such as the mass gap, as in
BCS superconductivity and charge density wave etc.
The key point is that these phenomena could be understood as an
instability from the Fermi liquid identified as the
fixed point of  the RG flow. They showed for the BCS case as an example
that when two incoming momenta sum to zero, the corresponding four-Fermi
interaction is {\it marginally relevant}
\footnote{Throughout this article,
we put in italic the terms  ``relevant,"
``marginal," and ``irrelevant" when used in the RG sense.}
in the RG sense, so that the interaction  causes
an instability that pushes away the system from the Fermi liquid.
The new insight gained in this approach is that one can
identify in a clear and simple way
the dynamics and kinematics that lead to the  phase transitions.
In this paper, we extend this approach to the condensation of negatively
charged kaons ($K^-$) in dense nuclear medium as in neutron stars
by considering the role of the
quadratic term in the effective potential entering in
kaon-nucleon  interactions. (The $K^+$ meson does not condense and the
neutral kaons $K^0$ and $\overline{K^0}$ are not relevant in neutron
stars.)

Recently Lee {\etal} \cite{rho1,rho2}
have shown by chiral perturbation theory ($\chi PT$) treated
to {\it in-medium} two-loop order (corresponding
to next-to-next-to leading order) that kaons can condense in dense
nuclear-star matter at a matter
density $\rho\lsim 4\,\rho_0$ where $\rho_0$ is the normal nuclear matter
density.
However to the order considered, many terms are
involved and it is not transparent which mechanism is in action for
triggering the condensation process.

In this note, we present a
renormalization group flow analysis to show what drives the process
of kaon condensation and in particular to indicate the basic mechanism
involved. The conclusion is that kaons must condense in s-wave, although
the analysis cannot give the critical density.

For the purpose of elucidating the basic concept, we find it sufficient to
study a toy model which we believe captures the essence of the physics
involved in kaon condensation. The corresponding action, $S$, can
be decomposed
into three parts: $S_K$ for the free kaon, $S_N$ for the nucleon and $S_{KN}$
for kaon-nucleon interactions. We assume
that nucleons  in nuclear matter are in Fermi-liquid state with the Fermi
energy $\mu_F$ and the Fermi momentum $k_F$. This state might arise from
chiral Lagrangians as some sort of ``Q-balls" or nontopological solitons.
(For a discussion on this, see ref.\cite{rho3}.)
For our purpose, it is crucial that the nuclear matter arises as a Fermi
liquid\cite{baym}. Defining  $\psi$ as  the nucleon field fluctuating around
the Fermi surface
such that the momentum integral has a cut-off $\Lambda_N$,
\be
k_F -\Lambda_N < |\vec{k}| < k_F + \Lambda_N,
\ee
the action in the nucleon sector can be written, schematically, in the
form
\be
S_N=
\int d\epsilon d^3k \psi^{\dagger}\left(\epsilon -\epsilon(\vec{k})\right)\psi
    +g \int (d\epsilon d^3k)^4 \psi^{\dagger}\psi^{\dagger}\psi\psi
    \delta^4(\epsilon, \vec{k})\label{naction}
\ee
where $\epsilon(\vec{k})$ is the nucleon energy measured as usual relative
to the chemical potential $\mu_F$ and the $\delta$ function assures the overall
energy-momentum conservation.
The second term with a coupling constant $g$ is a generic residual
Landau-Migdal four-Fermi (nucleon) interaction\footnote{Because of
the flavor degree of freedom, there are in general four independent four-Fermi
interactions for nuclear interactions. They are all {\it marginal}.}
which is {\it marginal}, so the normal nuclear matter is a fixed point,
which will be called ``Fermi-liquid fixed point" as in condensed
matter physics.
This is an immediate application of the
recent idea developed by Shankar\cite{shankar} and
Polchinski\cite{polchinski1,polchinski2} to nuclear matter.

We focus on kaons that are in chemical equilibrium with electron gas
in high density. So the  kaon chemical potential -- denoted $\mu_K$ -- is
the same as that of the electron. This follows from charge conservation.
In this situation, it is more convenient to
measure the kaon frequency relative to the kaon chemical potential.
Furthermore the kaon fluctuating momentum, $\vec{q}$,
is much smaller ({\ie}, in s-wave) than the kaon chemical potential,
so the kaon field $K$ can be taken to be essentially non-relativistic.
(This is most likely to be valid for kaon condensation in compact-star
matter where the kaon chemical potential is substantial but may not be
valid in dense
symmetric nuclear matter appropriate for heavy-ion collisions.)

Therefore  what is relevant is the small kaon energy fluctuation around
the chemical potential.
These properties of the kaon can be simply incorporated
by redefining the kaon field as
$ K \rightarrow \Phi e^{-i\mu_K  t}/\sqrt{2\mu_K}.$
The action for the kaon is
\be
S_K= \int d\omega  d^3q \Phi^{*}(\omega, \vec{q})\left(\omega -
q^2/2\mu_K\right)\Phi(\omega, \vec{q})-\int d\omega  d^3q\,\tilde{M}
 \Phi^*\Phi \label{kaction}
\ee
which follows from a relativistic Lagrangian by keeping terms of order
$1/\mu_K$.
The quadratic coupling(``mass''), $\tilde{M}$, in the second term is
$\frac{M_K^2 -\mu_K^2}{2\mu_K}$.
As with all effective
theories, we need a cut-off for the fluctuation. The momentum
cut-off for the kaon is defined by $\Lambda_K$, $ |\vec{q}| < \Lambda_K$,
which cannot be larger than $\mu_K$.
For simplicity, we shall
not distinguish  between
$\Lambda_K$ and $\Lambda_N$ and denote them generically $\Lambda$.

For kaon-nucleon interactions, we focus on s-wave kaons. (P-wave
kaons do not figure directly in condensation phenomena for the same
reason as the absence of p-wave pion condensation as mentioned below.)
The s-wave kaon-nucleon interaction involved
can be
summarized in a simple form
\be
S_{KN}= \int (d\omega d^3q)^2 (d\epsilon d^3k)^2 h  \Phi^*\Phi
\psi^{\dagger}
 \psi \delta^4(\omega,\epsilon, \vec{q}, \vec{k}).\label{knaction}
\ee
The important quantity here is the coefficient $h$. In chiral
Lagrangians, this
coefficient subsumes several terms of varying
importance, including energy dependence.  For instance, terms of higher chiral
order but quadratic in the nucleon field also take this form.
Specifically in chiral perturbation theory\cite{rho1,rho2}\footnote{
Our strategy in applying effective chiral Lagrangians to our many-body
problem is as follows: We assume
that we first retain suitable leading chiral order terms corresponding
to a mode elimination in the ``elementary-particle" space and then apply
the mode elimination around the Fermi sea {\it afterwards}.},
the $h$ includes\footnote{The s-wave $K^-p\Lambda (1405)$ coupling that
plays a crucial role in the threshold properties of kaon-nucleon interactions
and also in K-mesic atoms is not included in the model we are considering.
It turns out to play a minor role (as an {\it irrelevant term})
in the RG-flow argument, lending a support to the similar finding
in chiral perturbation theory of \cite{rho1,rho2}.}
the leading chiral order term [$O(Q)$] of the form
$\sim \frac{\omega}{f_\pi^2} K^\dagger K N^\dagger N$ and
the next-to-leading-order
terms [$O(Q^2)$], the so-called sigma term, $\sim \frac{\Sigma}{f_\pi^2}
K^\dagger K \bar{N} N$ and a higher derivative term,
$\sim \omega^2 K^\dagger K \bar{N} N$. (Here $K$ is the doublet kaon field,
$N$ the doublet nucleon field and $\omega$ is the kaon frequency that equals
$\mu_K$ at condensation.) In the case of s-wave kaon
condensation, what we are interested in is the flow of the quadratic term
({\ie}, the ``mass'' term) in the effective potential under the RG
transformation.

Now we have a well defined system of strongly interacting kaons and nucleons
as a system with a possible ground state configuration
determined by the chemical potentials of the kaon and the nucleon. This
configuration is supposed to result from many-body effects of the
nuclear matter. The interactions between fluctuating fields around
the ground-state configurations define the toy model, the action of which
is taken to be of the form
\be
S&=& S_K + S_{NK} + S_N \nonumber\\
 &=&  \int d\omega  d^3q \Phi^{*}(\omega, \vec{q})\left(\omega -
q^2/2\mu_K\right)\Phi(\omega, \vec{q})-\int d\omega  d^3q\,\tilde{M}
 \Phi^*\Phi \nonumber\\
&& + \int (d\omega d^3q)^2 (d\epsilon d^3k)^2 h \Phi^*\Phi \psi^{\dagger}
 \psi \delta^4(\omega,\epsilon, \vec{q}, \vec{k})\nonumber\\
&& + \int d\epsilon d^3k \psi^{\dagger}\left(\epsilon -\epsilon(k))\right)\psi
    +g \int (d\epsilon d^3k)^4 \psi^{\dagger}\psi^{\dagger}\psi\psi
    \delta^4(\epsilon, \vec{k}).
\label{toy1}
\ee

We now perform the RG analysis following
Shankar\cite{shankar} and Polchinski\cite{polchinski1,polchinski2}.
The low-energy effective theory with a cut-off $s\Lambda$ with $s < 1$ can
be obtained by integrating out the high frequency modes. The stability of
the system can then be determined by computing the
RG flow of the coupling constants of the interaction terms following this
mode-elimination procedure.

The scaling law of the kaon field is {\it defined} by requiring the kinetic
term of eq.(\ref{toy1}) to be invariant under scaling
$\omega \to s\omega$  and  $q \rightarrow \sqrt{s}q$, which is imposed
to make  $\omega$ and $q^2/2\mu_K$ scale in the same way.
Since the $\mu_K$ is fixed by the interaction
with the electron, we can keep it scale-free.
The scaling dimension of $\Phi$ then comes out to be\footnote{$[{{\cal O}}]$
stands for the scaling dimension of the object ${{\cal O}}$.} $[\Phi] = -7/4$
which follows simply from the scaling of the integration measure
$ [d\omega d^3q]= 5/2$ and the invariance condition
$[d\omega d^3q] + [\omega] + 2 \times [\Phi] =0.$
The second term (``mass'' term) is not invariant but {\it relevant},
$[d\omega d^3q] + 2 \times [\Phi] = -1,$
which means that the ``mass'' term grows as the fast modes are integrated
out. But this is not the entire story since there is a contribution from
the interaction term when fast nucleon modes in $s\Lambda<k<\Lambda$
are eliminated. To see what that does, we have to determine
the scaling behavior of the nucleon system.
We take the well-known procedure of the
``scaling toward the Fermi surface" where  only
the component perpendicular to the Fermi surface is scaled.
We have the scaling
$[d\epsilon d^3\vec{k}] = 2, \,\, [\delta^4(\epsilon, \vec{k})] = -2.$
Again keeping the kinetic part -- the fourth term of eq.(\ref{toy1})
-- invariant, we get the scaling rule for the nucleon,
$[\psi] = -3/2, \ \ [d\epsilon d^3\vec{k} \psi]= 1/2.$
This allows us to calculate the scaling of the third term of eq.(\ref{toy1})
$$2 \times([d\omega d^3 \vec{q} \Phi] + [d\epsilon d^3\vec{k}\psi])
+ [\delta(\omega, \epsilon)] + [\delta(\vec{q}, \vec{k})]
= 1/2.\label{third1}
$$
This shows that the kaon-nucleon four-point interaction is generally
{\it irrelevant}. The question we wish to address now is whether the scaling
law can be modified
by radiative (loop) corrections of the sort discussed recently by
Polchinski\cite{polchinski2} and Nayak and Wilczek\cite{wilczek}.
A simple analysis shows that up to one loop, there are no corrections that
change the tree-level scaling behavior. For example, the correction from
Fig. 1b is {\it irrelevant}
 since the overlapping kinematical region for the kaon and
nucleon shrinks as $s$ decreases.
This completes the counting of the naive scaling dimensions of the
terms involved.  At this point one might conclude that the common wisdom for
s-wave kaon condensation cannot be supported by the RG-flow
argument, since the attractive interactions involving
$\Sigma$ term which are supposed to drive the kaon
condensation are found
to be {\it irrelevant}.  However we will show that this {\it irrelevant} but
attractive term is what triggers the kaon condensation.

 Under the RG transformation with $\Lambda \rightarrow s\Lambda$, the
perturbation on the ``mass'' can be calculated as
\be
\delta \tilde{M} = s^{-1}\left(\tilde{M} -(1-s) \frac{\Lambda
h}{6\mu \pi^2} (3 k_F^2 + \Lambda^2(1+s + s^2)\right) -\tilde{M},\label{delm}
\ee
where the factor $s^{-1}$ comes from the scaling property of $\Phi$.
The one-loop RG equations can be obtained by putting $s = 1 -
\delta t$ in eq. (\ref{delm})
\be
{d\tilde{M}\over dt}=\tilde{M}-D h,\label{flow1}
\ee
\be
{dh\over dt}=-a h - A h^2 ,\label{flow2}
\ee
where, $D=\frac{3(1+\alpha^2)\alpha}{2\mu_K}\rho$,
$t=-\ln s$, $\alpha= \Lambda/k_F$. The constant $a$ can be obtained from
the scaling properties of the interaction terms: For the
interaction in eq.(\ref{knaction}), $a=1/2$.  The second term on the
right-hand side of  eq. (\ref{flow1}) is given by the diagram of Fig. 1a.
One can show that $A$ in eq. (\ref{flow2}) which can be
calculated from the diagrams in Fig.1b vanishes~\footnote{Here we assume
that the BCS-type instability for four nucleon
coupling is suppressed. In fact there is no
BCS-type instability involving strangeness flavor.}.
The factor $\Lambda\over k_F$ in the coefficient $D$ is due to the
restriction imposed on the momentum region, {\ie}, the momentum cut-off. We
are considering fluctuations near the Fermi surface.
The solution for the $\tilde{M}$-flow is given by
\be
\tilde{M}(t)=(\tilde{M}_0-{Dh_0\over 1+a})e^{t} + {Dh_0\over 1+a}e^{-at}
\label{sol}
\ee
with
\be
h(t)=h_0 e^{-at},\ \ \ h_0\geq 0.
\ee
In this formula, ${\tilde M}_0 $ must be positive as
the boson chemical potential cannot exceed its mass.

In order to understand the physics involved, we describe the flow for
general $a$ and $D$ in the $({\tilde M},h)$ plane.
For the moment, we shall forget that these quantities are fixed
in our model and allow arbitrary values to $a$ and $D$. In general,
$D$ is positive if the interaction is attractive,
negative if repulsive and zero if there is no interaction.
Also $a > 0$ if the interaction is {\it irrelevant},
$< 0$ if {\it relevant} and $=0$ if {\it marginal}. We describe the flow of
each case in Figure 2. If there were no
attractive interaction $h$  giving rise to a density-dependent
term in the mass flow  [eq.(\ref{flow1})], then the flow would have gone
straight up in the mass direction in the $ (\tilde{M}, h)$ plane if we start
from any initial value $\tilde{M}_0>0, h_0>0$. See Fig. 2 a,b,d,f,h.
In this case, if the mass is positive at some scale, then
it will remain positive at all scale, so there would be no region
 from which the mass could flow to negative direction.
However, with inclusion of an {\em attractive} interaction,
the mass flow becomes qualitatively different. See Fig. 2 c,e,g.
Our case defined by the toy model (\ref{toy1}) corresponds to Fig. 2c.
This analysis shows us the nontrivial aspects of {\it irrelevant}
interactions in determining the direction of ``mass" flow.
We also note that the Gaussian fixed point is a saddle point
of the RG flow and unstable.

The interesting feature of this flow in connection to kaon
condensation is that there are two types
of flows depending upon the sign of the coefficient of $e^t$ in eq.(\ref{sol}).
For
\be
D < \frac{\tilde{M}_0}{h_0}(1+a),
\ee
it flows toward $+\infty$ whereas for $D > \frac{\tilde{M}_0}{h_0}(1+a)$,
it flows downward to $-\infty$.
The flow eventually crosses the zero ``mass" axis with the speed
of the RG flow determined by the density $\rho_N$.
Thus in the low-energy limit, the sign of the ``mass" term becomes
negative if the
the initial point belongs to the shaded region in Fig. 2c. This  signals a
 meson condensation as we explain below.

In the mean-field approach, when the effective mass of the kaon decreases
and approaches $\mu_K$, kaon condensates will develop. This may be expected
from the ideal bose gas  distribution
\be
n(k)={1\over e^{\beta[\epsilon(k)-\mu]}-1},\label{bed}
\ee
where the ground-state occupation number diverges when
the chemical potential equals the ground-state energy.
What the RG analysis tells us is that
this tendency for the kaon condensation will be eliminated
in the low energy limit
if the ``mass" flows toward a point $(\infty,0)$ in the $(\tilde{M},h)$
plane. Therefore the flow toward
a point $(-\infty,0)$ or crossing the zero ``mass" axis can be
identified as a signal for kaon condensation in the RG analysis.
A similar observation can be made from the mechanism of spontaneous
symmetry breaking.
As mass becomes negative, the vacuum becomes unstable, and something must
happen for the system to restore the stability.
What happens is that the system develops
a positive $\Phi^4$ term for stability and the vacuum gets shifted to
a new one with non-zero VEV, $<\Phi>\ne0$. This is the familiar mechanism for
spontaneous symmetry breaking induced by the sign change in the
mass. Therefore the ``mass" flow crossing zero can be naturally
associated to an instability conducive to a certain phase transition.
{\it The key point of this paper is that the downward flow of Fig.2c can
be associated with the instability leading to kaon condensation.}

An interesting point to note in the above analysis is that
the main driving term for $m$ in the RG equation is
provided by the {\it relevant} term, {\ie}, the mass term,  but
the direction of the flow is determined by the {\it irrelevant}
interaction term. An
illustrative application of this observation is that in the
chiral limit, where
there is no mass term, condensation cannot be driven
by the remaining {\it irrelevant} interactions.
This is consistent with
the fact that, in the chiral limit, chiral symmetry prevents the flow
of the mass and there cannot be any phase transition.

To  understand better the role played by chiral symmetry breaking
in kaon condensation, we examine how chirally symmetric terms
and symmetry-breaking terms differ in their contribution to the mass or energy.
To do this, we take one term each from three classes of kaon-nucleon
interactions:
(i) a mass term, (ii) an explicit chiral-symmetry-breaking interaction term
({\ie}, the  $\Sigma$ term), $\frac{\Sigma}{f_{\pi}^2}K^{\dagger}K\bar{N}N$,
and (iii) a chiral-symmetry-preserving interaction term,
$ c\partial_{\mu}K^{\dagger}\partial^{\mu}K \bar{N}N$. Notice that
in our toy model above, (i) and (ii) are included in the $h$ term.
With one-loop corrections to the self-energy,
the relativistic inverse propagator becomes
\be
G^{-1}(k,\omega)=\omega^2 - k^2 -m^2 -\Pi(\omega,k) \label{propa}
\ee
where $m$ is the meson mass,
$\Pi=C(\omega^2 - k^2)-H $ with $C$ a constant proportional to $c$ and
$H=3\frac{\Sigma}{f_{\pi}^2}(1+(\Lambda/k_F)^2)(\Lambda/k_F)\rho_N $
is the contribution from the $\Sigma$ term.  Equation (\ref{propa}) can
be rewritten as
\be
G^{-1}(k,\omega)=(1-C)(\omega^2 - k^2-{\tilde m}^2),
\ee
where  $1-C$ is the wave function renormalization factor and
$\tilde{m}^2=(1-C)^{-1}(m^2-H).$
The mass shift is
\be
\delta m^2=(1-C)^{-1}(m^2-H) -m^2=(1-C)^{-1} (C m^2 -H) .\label{dmass}
\ee
Let us first turn off the $\Sigma$ term.
Then the mass correction is simply proportional to the original mass,
$\delta m^2= \frac{C}{1-C}  m^2 .$
Thus if the original mass is small, then the
chirally symmetric interaction gives a small mass correction. This is
just the precise statement of PCAC.
If there were no mass term to start with,  of course,
there would be no mass correction. In this case there would be no
$\Sigma$ term either.
The Goldstone boson mass is protected by chiral symmetry.
However if the symmetry breaking is not small, chirally symmetric
interactions can contribute importantly to mass or energy.
The above analysis shows that repulsive interaction
$(C>0)$ increases the meson mass while attraction decreases it.
When the $\Sigma$ term is present, the $\Sigma$ term attraction and
the $C$ term repulsion can compensate each other.
For pion-nucleon interactions,
the two terms effectively cancel, leaving the pion mass more or less
unchanged, while in the case of kaons, the $\Sigma$ term attraction
wins out.

An exactly parallel RG-flow analysis could be made for s-wave {\it pion}
condensation.
As in the case of the kaon, it is the mass term that can cause
instability with the $\Sigma_{\pi N}$ term determining the direction
of the ``mass" flow. We therefore expect a similar mass flow as in Fig. 2c,
leading to s-wave condensation. However the pion is {\it almost}
massless on the strong-interaction scale and the chiral symmetry
{\it almost} protects its mass. Any small repulsion would counterbalance
the small attraction associated with the $\Sigma_{\pi N}$ term. Thus
as we know from on-shell charge-symmetric $\pi N$ and $\pi$-nuclear
amplitudes, the repulsive term $\sim \omega^2$ -- which includes,
in nuclear medium,
the well-known Pauli exclusion principle effect\cite{nambu} --
overpowers the $\Sigma_{\pi N}$ attraction, preventing
the pion mass from going to zero. More specifically,
in the RG analysis, $\tilde{M}_0 -Dh
\geq 0$ for the s-wave pion due to the repulsive term $\sim
\omega^2$ and there is no instability
toward s-wave pion condensation
This is consistent with PCAC which says that the soft-pion point --
which is physical in the chiral limit -- and the would-be condensation
point are smoothly interpolated and hence the absence of s-wave
pion condensation in the chiral limit implies its absence at any
off-shell point. This is
also consistent with the observations made in lattice
gauge calculations
(in high temperature) and QCD sum rule calculations which indicate that the
pion mass does not change appreciably as temperature or density is increased.
Perhaps the most significant point to note is that
that in two extreme limits, the chiral limit
and the heavy-quark limit, bose condensations cannot take place and that
kaons can condense at low enough density
because the kaon is neither very light nor very heavy.

The p-wave pion condensation studied extensively since many years
is different from the s-wave case from the point of view of the RG flow.
As shown, in the case of the s-wave condensation, it is the
fluctuation in the ``mass" direction around zero meson three-momentum
that affects crucially
the low-energy dynamics (the role of the s-wave pion interaction
is minor in the sense that
the interaction itself is not responsible for the instability).
The condensate so developed is spatially uniform, with the order parameter
being space-independent.
However the p-wave condensation involves non-zero spatial momentum,
with the condensate varying in space. For this, Yukawa interactions with
the meson of non-zero momentum are required. For an instability to
set in, a {\it relevant} interaction
in the appropriate channel would be necessary.
However, we can show, following  Polchinski \cite{polchinski2}, that
the $\pi NN$ Yukawa interaction, when radiatively corrected, becomes
{\it irrelevant} and the four-Fermi interaction in the pion channel is
at most {\it marginal}. Therefore p-wave pions do not condense.
It is not clear how this reasoning is related to the mean-field argument
\cite{baym2} where the strength of the Landau-Migdal interaction $g_0^\prime$
plays a crucial role in pushing up the critical density.

One might ask what the effect of the kaon condensation is on the fermion
(nuclear) system. The fermion system involved here
is a bit intricate because of the Fermi surface.
Since the density that gives the Fermi momentum is a fixed and
physically controlled quantity, we must
compensate the change by introducing a counter term\cite{shankar}. This means
that the flow of the Fermi surface should be prevented by {\it fiat}. This
is the basic difference between the kaon and nucleon mass-like
terms. While the
flow of the kaon mass is allowed and leads to instability,
we cannot allow the flow of the Fermi surface since the density must be fixed.

Now once the kaon condensation takes place, one can
decompose the kaon field into a condensed part and a fluctuating part.
Focusing on the s-wave condensation only, we can write
\be
K(x,t)=K_0 e^{-i\mu t} +\int {d^4 k\over (2\pi)^4 } \Phi(k)e^{ik\cdot x}
\label{division}
\ee
where $k=(\omega,{\bf k})$ and $K_0$ represents the condensed part.
With eq.(\ref{division}), the interaction term in the model Lagrangian,
eq.(\ref{toy1}), contains an additional term
\be
\int d\epsilon d^3k h |K_0|^2 \psi^{\dagger}\psi .
\label{toy2}
\ee
This means that the net effect of the kaon condensation on fermion dynamics
is to change the fermion chemical potential by $\mu_F \to \mu_F + h|K_0|^2$.
If we fix the density of the nucleons, then the Fermi surface is fixed,
$k_F \sim n^{1/3}$. The effective fermion mass must therefore decrease
since $\mu_F=k_F^2/2m^*_F$ must increase.

In this note we have discussed how the RG approach can be
used for strongly interacting hadronic matter, {\eg}, for the case where
kaon condensation might take place.  The flow of the kaon
``mass" ({\ie}, the quadratic coupling) is
argued to be a signature for the instability of the kaon-nuclear system.
We interpret the RG flow across the zero mass axis as the instability
that leads to the s-wave kaon condensation.
An analogy to the familiar spontaneous symmetry breaking (SSB) mechanism has
 been used, where the negative mass term flows
due to the mode elimination from the $\Phi^4$ term and eventually crosses
zero. At this point
the vacuum changes into the ``true" vacuum with $<\Phi> \neq 0$.
We should emphasize that the mechanism we are
suggesting is so general that it is independent of the detailed structure
and of the number of interaction terms.
What we are showing here is that in {\it any}
interacting boson-fermion systems,
s-wave bose condensation must take place if there is an attractive interaction
between fermion and {\it massive} boson and if the density is high enough.
Similar RG flow analyses can
be made for cases where there are more than two coupling constants.
Although our discussion is based on a simplified model, it contains,
however, enough generic feature that encapsulates the basic physics
implied by the underlying symmetry of the system, namely, chiral symmetry
and its breaking\cite{rho3}.

\subsection*{Acknowledgments}
\indent We are grateful for helpful discussions with Chang-Hwan Lee.
Two of us (HKL and SJS) would like to acknowledge
the hospitality of
Service de Physique Th\'eorique of CEA Saclay where part of this
work was done.  This work is supported in part by Ministry of
Education (BSRI-94-2441) and in part by Korea Science and
Engineering Foundation(94-1400-04-01-3).


\newpage
\vskip1cm
\centerline{\bf  FIGURE CAPTIONS}

\vskip 1cm
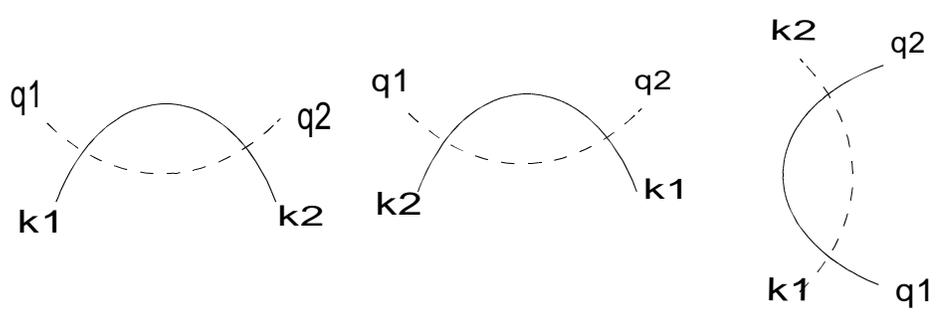
\noindent {\bf Figure 1:} One-loop correstions.

 a) Mass correction (to $\tilde{M}$) due to
$ K^\dagger KN^\dagger N$-type terms;

 b) One-loop coupling constant correction (to $h$)
due to $K^\dagger KN^\dagger N$-type terms.

The solid lines represent nucleons and the dotted lines
represent mesons.
\vskip 1cm
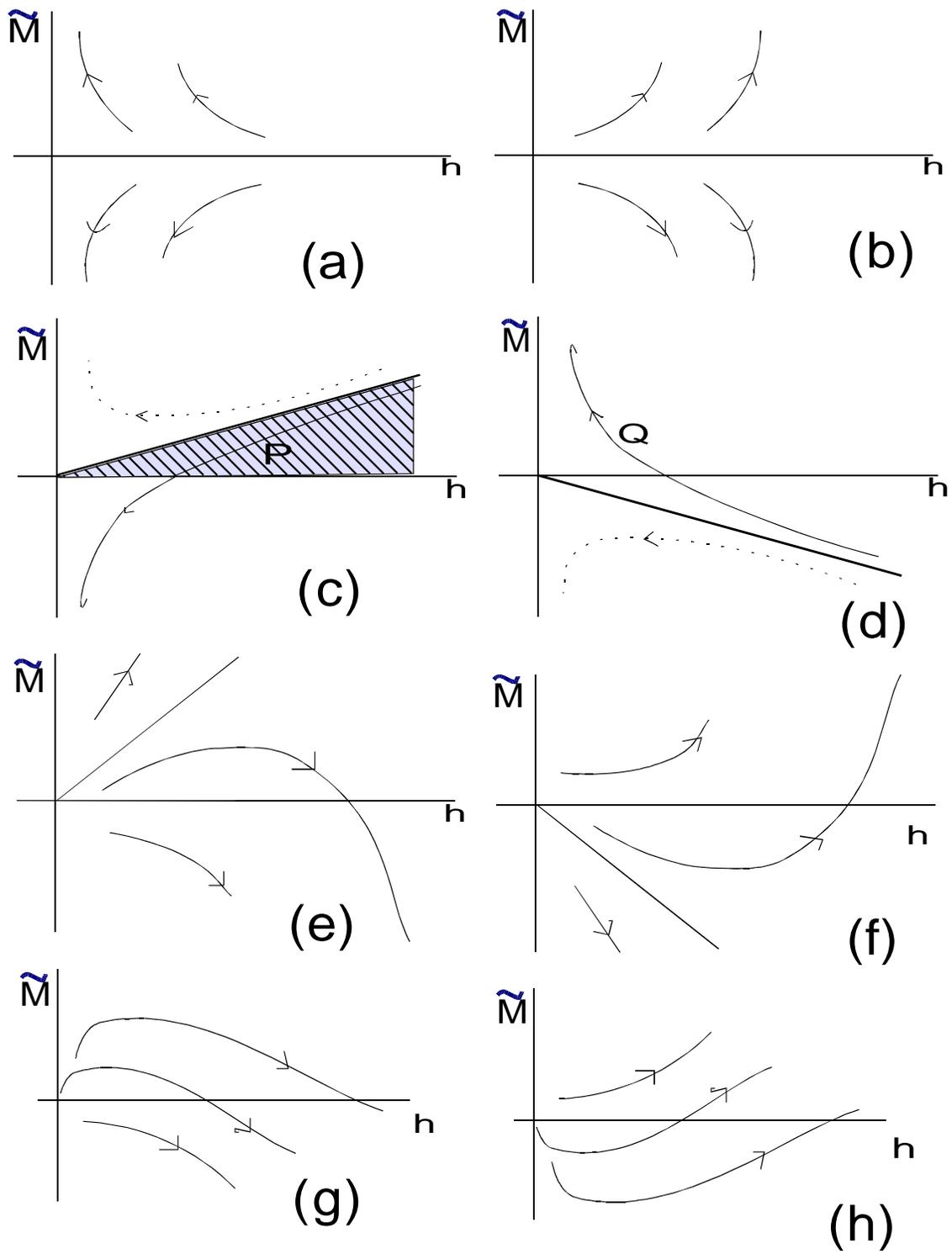
\noindent {\bf Figure 2:} Possible flows for given values of $(a,D)$ in the
$(\tilde{M},h)$ plane.

a,b) $D=0$, no coupling; a) $h$ is {\it irrelevant} and b) {\it relevant}.
No possibility for sign change of $\tilde{M}$.

c) $D>0$ and $a>0$; the interaction term is attractive and {\it irrelevant}.
Sign change of $\tilde{M}$ is inevitable for {\it some} $(\tilde{M}_0,h_0)$.
This is the case that corresponds to the situation encountered in
the chiral perturbation theory calculation of refs.\cite{rho1,rho2}.

d) $D<0$ and $a>0$; the interaction term is repulsive and {\it irrelevant}.
 Sign change of $\tilde{M}>0$ is impossible.

e) $D>0$ and $-1<a<0$; the interaction term is attractive and weakly {\it
 relevant}. Sign change is inevitable for {\it some} $(\tilde{M}_0,h_0)$.

f)  $D<0$ and $a<0$; the interaction term is repulsive and {\it relevant}.
   Sign change is impossible for {\it any} $\tilde{M}_0>0$

g) $D>0$ and $a<-1$; the interaction term is attractive and {\it relevant}.
   Sign change is inevitable  for {\it any} $\tilde{M}_0>0$

h) $D<0$ and $a<0$; the interaction term is repulsive and {\it relevant}.
   Sign change is impossible for {\it any} $\tilde{M}_0>0$

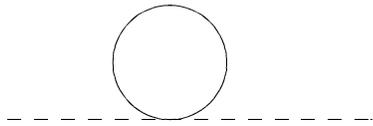
\end{document}